\documentclass{article}                    
\usepackage{graphicx}
\usepackage{amssymb}
\usepackage{amsmath}

\begin{document}

\title{Regime variance testing  - a quantile approach}



\author{\textbf{Janusz Gajda,      Grzegorz Sikora,        Agnieszka Wy{\l}oma{\'n}ska}\\\\
        Hugo Steinhaus Center, Institute of Mathematics and Computer Science\\
  Wroclaw University of Technology, Poland\\
   Tel.: +48-71-320-31-83\\
   janusz.gajda@pwr.wroc.pl, grzegorz.sikora@pwr.wroc.pl\\ agnieszka.wylomanska@pwr.wroc.pl}  
          

\date{\today}

\maketitle

\begin{abstract}
This paper is devoted to testing time series that exhibit behavior related to two or more regimes with different statistical properties.  Motivation of our study are two real data sets from plasma physics with observable two-regimes structure. In this paper we develop estimation procedure for critical point of division the structure change of a  time series. Moreover we propose three tests for recognition such specific behavior. The presented methodology is based on the empirical second moment  and its main advantage is lack of the distribution assumption. Moreover, the examined statistical properties we express in the language of empirical quantiles of the squared data therefore the methodology is an extension of the  approach known from the literature \cite{structural2,structural3,structural4,structural5}.  The theoretical results we confirm by simulations and analysis of real data of turbulent laboratory plasma.
\end{abstract}

\section{Introduction}\label{introduction}
The main issue in real data analysis is testing distribution. This problem appears not only in   case of independent identically distributed (i.i.d.) sample \cite{dist1,dist2,dist3} but also when we calibrate a model to real data set \cite{model1,model2,model3}. In this case the distribution is fitted to the residual series that is assumed to be i.i.d. But many independent variables seem to display changes in the underlying data generating process over time \cite{structural1} therefore they can not be considered as identically distributed sample.  This typical behavior we observe also in time series described in Section \ref{moti} that presents increments of floating potential fluctuations  of turbulent laboratory plasma for the  small torus radial position $r=37$ cm. For this data set  the known statistical tests  for stationarity mentioned in Section \ref{econom} \cite{bib:DickeyFuller1979,bib:DickeyFuller1981,bib:SaidDickey1984,bib:PhillipsPerron1988,bib:KPSS1992} are not useful.  What  more, under some assumptions they indicate the data are i.i.d. that is in contradiction with behavior of the data observable in Figure \ref{time_series_1}.

Therefore in this paper we introduce three tests that can be useful to time series for which we observe more than one regime with different statistical properties. Two of them are  visual therefore we call them pre-tests and  propose to use in the preliminary analysis to identify the specific behavior. In order to confirm two or more regimes in the data set we have developed  statistical test for regime variance. Moreover, we also introduce the estimation procedure for the critical point  that divides examined time series into two parts with different statistical properties.  However, the inspection only of the data can lead sometimes to the wrong preliminary choice of the model therefore the mentioned tests are based on  behavior of the empirical second moment of the examined time series.   The advantage of the methodology based on the empirical second moments is emphasized in \cite{nasza,moja} and is also confirmed by the bottom panel of Figure\ref{time_series_1} that presents the squared data for which the difference between two regimes is more visible. In the  presented methodology we do not assume the distribution because the introduced tests exploit only the empirical properties of examined data set. What is more important, they can be used to data for which the point of division into two regimes is well-defined (is clearly observable), but also for data for which the point is not visible. Moreover, we show by simulation study that the proposed methodology can be also useful for infinite-variance time series. 
 
The rest of the paper is organized as follows: in Section \ref{moti} we present the examined data sets that are motivation of developing the presented methodology.  Next, in Section \ref{econom} we overview the known statistical tests for stationarity and present the estimation procedure of recognition the critical point introduced in \cite{bib:Tsay1988}.  In Section \ref{visual} we introduce two visual pre-tests that indicate at specific behavior of examined time series, i.e. two regimes related to different statistical properties. In this Section we propose also the innovative procedure of estimation for critical point based on the behavior of empirical second moment of real data set and present the simulation study.  In Section \ref{test} we introduce the statistical method for testing regime variance and test the procedure by using simulated data. In the next Section we analyze real data sets form plasma physics in the context of presented methodology.  Finally, the last Section  gives a few concluding remarks.
\section{Motivation}\label{moti}
\begin{figure}[htb]
\centering \includegraphics[height=6cm]{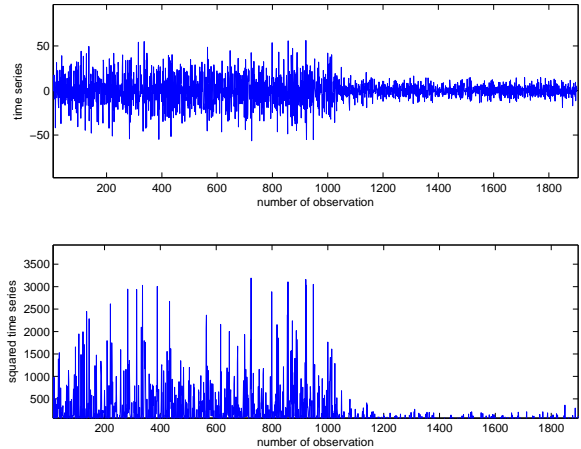}
\caption{The  empirical time series from plasma physics that presents increments of floating
potential fluctuations (in volts) of turbulent laboratory plasma for the  small torus radial position $r=37$ cm (top panel) and squared time series (bottom panel). }\label{time_series_1}
\end{figure}

Motivation of our study is presented in Figure \ref{time_series_1} real data set. This time series describes increments of floating
potential fluctuations (in volts) of turbulent
laboratory plasma for the  small torus radial position $r=37$ cm. Precise
description of the  experiment is presented in \cite{chec2}. The signal was registered on 15 June 2006 with
movable probe in scrape-off layer (SOL) plasma of stellarator
"URAGAN 3M". Because the signal was registered every $0.0000016$ second therefore total length of time series is $30000$, but to the analysis we only take $1900$ observations between $12000-13900$.  
As we observe in Figure \ref{time_series_1} the empirical data set exhibits very special behavior, namely the statistical properties of the time series change in time. It can be related to the fact that the first observations constitute random sample that comes from another distribution, than the last part or  those two parts come from the same distribution with different parameters. Therefore we can suspect that the  time series  satisfies the following property:
\begin{eqnarray}\label{data}
X_i\stackrel{d}{=}
\left\{
\begin{array}{ll} 
X& \mbox{for $i\leq l$}\\
&\\
Y & \mbox{for $i> l$,}
\end{array}
\right.
\end{eqnarray}
where $X$ and $Y$ are independent and have different statistical properties and $l$ is fixed point. As we have mentioned in Section \ref{introduction}, inspection of the data can lead sometimes to the
wrong preliminary conclusions, therefore we propose to consider the  squared time series. As we observe in Figure \ref{time_series_1}, the difference between two parts is more visible for squared data.  The statistical properties  we express in the language of quantiles of squared time series and  we assume the random variables $X^2$ and $Y^2$ in relation (\ref{data}) have  different quantiles $q_{\alpha/2}$ and $q_{1-\alpha/2}$ for given confidence level $\alpha$. Here we take the notation $q_{a}$  as quantile of order $a$. 

\begin{figure}[htb]
\centering \includegraphics[height=6cm]{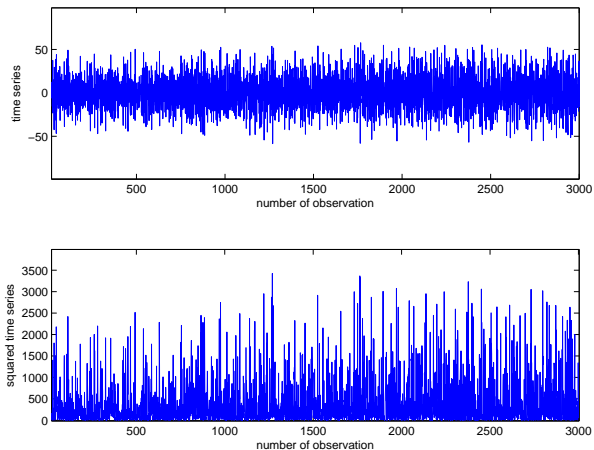}
\caption{The  empirical time series from plasma physics that presents increments of floating
potential fluctuations (in volts) of turbulent laboratory plasma for the  small torus radial position $r=36$ cm (top panel) and squared time series (bottom panel). }\label{time_series_2}
\end{figure}

After preliminary analysis of the data set and confirmation that it constitutes realizations of independent random variables (see Figure \ref{acf_time_series_1}) we have tested hypothesis of the same distribution of time series. The known statistical tests such as Augmented Dickey-Fuller, Phillips-Perron or Kwiatkowski--Phillips--Schmidt--Shin test for stationarity  reject the hypothesis that the data are nonstationary (in the sense presented in Section \ref{econom}) that suggests they are not useful for this data set. Therefore  we propose three tests that can be used for data that exhibit similar behavior as this observed in Figure \ref{time_series_1}, but also to this that after preliminary analysis we can not reject the hypothesis about the same distribution. An example is shown in Figure \ref{time_series_2}.  This time series presents increments of floating
potential fluctuations (in volts) of turbulent
laboratory plasma for the  small torus radial position $r=36$ cm. Similar as for the first data set, the signal was registered on 15 June 2006 and total number of observations was $30000$ but to illustration we take only observations from $12000$ to $15000$. After analysis of the time plot for series and squared series we can suspect that the data can not be considered as identically distributed sample but here the point, when some statistical properties change is not so visible as for the first data set.   Moreover the mentioned tests for stationarity presented in Section \ref{econom} indicate that under some assumptions the time series can be considered as stationary process but in next Sections we will show that this hypothesis is not true. Moreover we will find such point that divides examined data into two i.i.d. samples.

\section{Statistical tests for stationarity}\label{econom}

In order to make any inferences about the structure of a time series we need some regularity over time in the behavior of the underlying series. This regularity one can formalize using a concept of stationarity, see \cite{bib:CryerChan}. We say that time series is weakly stationary if the mean of the series is constant over time and the covariance between observations on time $t$ and $s$ depends only on their absolute difference $|s-t|$. 

However stationarity is not a common feature of time series and mostly we observe nonstationary behavior of the process. There are several types of nonstationarity. The trend nonstationarity means that the data posses some deterministic trend (for example  linear trend) but otherwise are stationary. This can be easily seen based on autocorrelation function (for instance linear trend can be seen as a linear slow in time decay of autocorrelation function) \cite{bib:CryerChan}. The second type of nonstationarity is called difference nonstationarity, which means the process  has to be differenced in order to become stationary. This two types examples of nonstationarity are often encountered in real-life data. The class of unit-root tests help to distinguish difference from trend nonstationarity. Under the null hypothesis that the series is difference nonstationary one can mention here Dickey-Fuller unit root tests  \cite{bib:DickeyFuller1979,bib:DickeyFuller1981,bib:SaidDickey1984} and Phillips-Perron unit root tests  \cite{bib:PhillipsPerron1988}. Testing in opposite direction, namely assuming that time series is trend stationary against it is difference one can apply the KPSS test due to Kwiatkowski, Phillips, Schmidt and Shin \cite{bib:KPSS1992}.

Mentioned above types of nonstationarity can be successfully tested and recognized from the data but they are not the only problems one may encounters during data analysis. Atypical observations, level shifts or variance change are common features of many real-life data sets \cite{structural1,structural2,structural3}. Neglecting such effects may lead to inaccurate estimation of parameters of the model and in consequence inaccurate or completely wrong prediction. In present work we will discuss the effect of variance change in the data sets, thus there is no trend and differenced data have the same behavior as initially before differentiation.  Such specific two-regimes time series was also considered in \cite{bib:Tsay1988}, where  the following model for the innovations (independent sample) was considered:
\begin{equation}
\label{Tsay:change}
\epsilon'_i = \left\{
\begin{array}{ll}
\epsilon_i & \text{if } i < l\\
\epsilon_i(1+\omega_V) & \text{if } i \geq l.
\end{array} \right.
\end{equation}
for some point $l$, fixed number $\omega_V$ and under the assumption $\{\epsilon_i\}_{i=1}^n$ constitutes i.i.d. random variables from normal distribution.  We can thus calculate the variance ratio of $\epsilon'_i$ before and after the structural change:
\begin{equation}
\label{Tsay:VarianceRatio}
\hat{r}_l=\frac{(l-1)\sum_{i=l}^n \epsilon'^2_i}{(n-l+1)\sum_{i=1}^{l-1} \epsilon'^2_i},
\end{equation}
where $(l-1)$ and $(n-l+1)$ are greater than zero. The variance ratio is an estimate of $(1+\omega_V)^2$ and is likelihood ratio test statistics of variance change under the assumption of normality. The test is the most powerful for step change in variance when the point $l$ is known. If the critical point is unknown one can apply variance ratio statistics to find it. In this case we compute the variance ratio statistics for stochastically independent series and obtain its minimum $\hat{r}_{min}$ and maximum $\hat{r}_{max}$ values:
\begin{equation*}
\begin{split}
\hat{r}_{min}=\min_{h\leq l \leq n-h}\{ \hat{r}_l\}, \\
\hat{r}_{max}=\max_{h\leq l \leq n-h}\{\hat{r}_l\},
\end{split}
\end{equation*}
where $h$ is the positive integer denoting the minimum number of observations used to estimate the variance at the beginning and at the end of the sample.  Then we calculate
\begin{equation*}
\hat{r}=\max\{\hat{r}_{min}^{-1},\hat{r}_{max}\}.
\end{equation*}
The critical point $l$ is the one at which $\hat{r}$ occurs.

The complete description of procedure for detecting and adjusting the time series with two-regimes structure of the type \eqref{Tsay:change} is presented in \cite{bib:Tsay1988}. Because the presented methodology is based on the assumption of normal distribution, that is a main disadvantage, therefore in the next Section we introduce the innovative procedure of estimation for the critical point that does not require any assumption of the distribution. This procedure is based on the behavior of empirical second moment of examined time series and is compatible with two visual pre-tests for two-regimes structure.

\section{Visual pre-tests for regime variance}\label{visual}

\begin{figure}[th!]
\label{figure1}
\centering
\includegraphics[height=8cm]{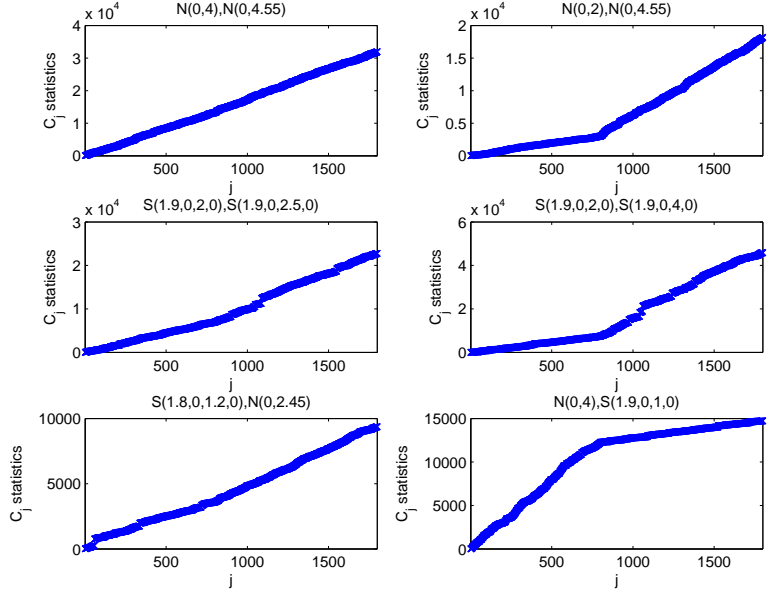}
\caption{The $C_j$ statistics defined in (\ref{Ck}) for two considered scenarios. On the left panel we demonstrate results for cases when the parameters of distributions are close to each other (the first scenario). The right panel presents  cases of distributions with very different parameters (the second scenario).}\label{visual_1}
\end{figure}

\begin{figure}[th!]
\label{figure1}
\centering
\includegraphics[height=8cm]{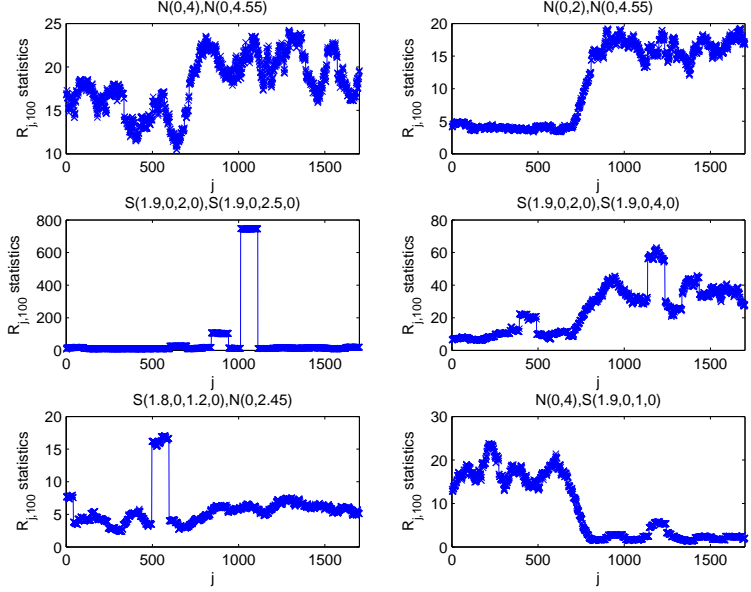}
\caption{The $R_{j,100}$ statistics defined in (\ref{R}) for two considered scenarios. On the left panel we demonstrate results for cases when the parameters of distributions are close to each other (the first scenario). The right panel presents  cases of distributions with very different parameters (the second scenario).}\label{visual_2}
\end{figure}

In the first part of this Section we present two visual pre-tests that can confirm information if the observed time series $X_1,X_2,...,X_n$ constitutes sample that satisfy relation (\ref{data}). Those two pre-tests are based on the behavior of empirical second moment of the data. In the first method we propose to consider the following statistics:
\begin{eqnarray}\label{Ck}
C_j=\sum_{i=1}^jX_i^2,~~j=1,2,...,n.
\end{eqnarray}
If the random variables  $X$ and $Y$ given in relation (\ref{data}) have distributions with finite second moments $\sigma_1^2$ and $\sigma_2^2$, respectively , then the statistics $C_j$ has the following property:
\begin{eqnarray}\label{data11}
E(C_j)=
\left\{
\begin{array}{ll} 
j\sigma_1^2& \mbox{for $j\leq l$}\\
j\sigma_2^2+l(\sigma_1^2-\sigma_2^2)&\mbox{for $j> l$.}
\end{array}
\right.
\end{eqnarray}
If $\sigma_1^2=\sigma_2^2$, then the mean of $C_j$ statistics is equal to $\sigma_1^2j$ for all $j=1,2,...,n$, therefore for i.i.d. sample expected value of the  statistics  is a linear function with the shift parameter equal to zero. Of course this relation is not satisfied for distributions with infinite variances, but even in this cases we observe significant changes in behavior of $C_j$ statistics. Results of this pre-test we present in Figure \ref{visual_1} for different distributions of random variables $X$ and $Y$ in relation (\ref{data}). We consider two cases each consisting of three distributions, namely pure Gaussian ($\mathcal{N}(\mu,\sigma)$), pure L\'evy--stable ($\mathcal{S}(\alpha,\beta,\sigma,\mu)$), and Gaussian--L\'evy--stable. In the first scenario we consider the case when the parameters of distributions are close to each other and thus the structure change point is not well visible in the visual pre-test (see left panel of Figure \ref{visual_1}). In the second scenario we consider distributions with very different parameters, thus the critical point is observable in the simulated sample (the right panel of Figure \ref{visual_1}). \\
For the first scenario we consider the following cases:
\begin{itemize}
\item the pure Gaussian case with $\mathcal{N}(0,4)$ and $\mathcal{N}(0,4.55)$ distribution for first 800 and last 1000 observations, respectively,
\item the pure L\'evy--stable case with $\mathcal{S}(1.9,0,2,0)$ and $\mathcal{S}(1.9,0,2.5,0)$ distribution for first 800 and last 1000 observations, respectively,
\item the L\'evy--stable--Gaussian case with $\mathcal{S}(1.8,0,1.2,0)$ and $\mathcal{N}(0,2.45)$  distribution for first 800 and last 1000 observations, respectively.
\end{itemize}
In the second scenario we consider following parameters of distributions:
\begin{itemize}
\label{close:parameters}
\item{the pure Gaussian case with $\mathcal{N}(0,2)$ and $\mathcal{N}(0,4)$ distribution for first 800 and last 1000 observations, respectively,}
\item{the pure L\'evy--stable case with $\mathcal{S}(1.9,0,2,0)$ and $\mathcal{S}(1.9,0,4,0)$ distribution for first 800 and last 1000 observations, respectively,}
\item{the Gaussian--L\'evy--stable case with $\mathcal{N}(0,4)$ and $\mathcal{S}(1.9,0,1,0)$  distribution for first 800 and last 1000 observations, respectively.}
\end{itemize}
In the second visual pre-test  we observe behavior of the empirical second moment of the data from windows of width $k>0$. The examined statistics has the following form:
\begin{eqnarray}\label{R}
R_{j,k}=\sum_{i=j+1}^{j+k}X_i^2,~~j=0,1,....n-k,
\end{eqnarray} 
where $k$ is a given positive number called window width. We assume $k<l$. For finite variance distributions of $X$ and $Y$ we can also calculate the expected value of $R_{j,k}$ statistic, namely:
  \begin{eqnarray}\label{data22}
E(R_{j,k})=
\left\{
\begin{array}{ll} 
k\sigma_1^2& \mbox{for $j+k\leq l$}\\
j(\sigma_2^2-\sigma_1^2)+l(\sigma_1^2-\sigma_2^2)+k\sigma_2^2&\mbox{for $j+1\leq l< j+k$}\\
k\sigma_2^2&\mbox{for $j+1> l$,}\end{array}
\right.
\end{eqnarray}
where $\sigma_1^2$ and $\sigma_2^2$ are the second moments of the random variables $X$ and $Y$ respectively. As we observe in (\ref{data22}), the mean of $R_{j,k}$ statistics for given window width is constant when $j\leq l-k$ or $j> l-1$. For $l-k< j\leq l-1$ the  statistics has mean that is a linear function with respect to $j$. When $X$ and $Y$  have the same distributions, then expected value of the statistics defined in (\ref{R}) is constant for given $k$. The results of this pre-test are presented in Figure \ref{visual_2} for two considered scenarios with different distributions presented above. 
\subsection{Estimation procedure for the critical point}
In this part we introduce the innovative  method of estimating the critical point of change the statistical properties in the sample that fulfills relation (\ref{data}). The idea of estimation procedure comes from the first visual pre-test described above. More precisely, we use the statistics $C_j,$ $j=1,2,\ldots,n$ defined in  (\ref{Ck}) and its mean function $E(C_j)$ given in (\ref{data11}).

The algorithm starts with dividing for fixed $k=1,2,\ldots,n$ the $C_j$ statistics into two sets $\left\{C_j: j=1,2,\ldots,k\right\}$ and $\left\{C_j: j=k+1,k+2,\ldots,n\right\}.$ Next, we fit the linear regression lines $y_j^1(k):=a_1(k)j+b_1(k)$ and $y_j^2(k):=a_2(k)j+b_2(k)$ to the first and the second set respectively. From ordinary regression theory for such lines 
the sums of distance squares $\sum_{j=1}^k{(C_j-y_j^1(k))^2}$ and $\sum_{j=k+1}^n{(C_j-y_j^2(k))^2}$ are minimized and therefore
the line coefficients have the form, see \cite{draper}:
\begin{eqnarray}\nonumber
a_1(k)=\frac{\sum_{j=1}^k{jC_j}-\frac{(k+1)}{2}\sum_{j=1}^k{C_j}}{-\frac{1}{4}k(k+1)^2+\frac{1}{6}k(k+1)(2k+1)},\quad b_1(k)=
\frac{\frac{1}{3}(2k+1)\sum_{j=1}^k{C_j}-\sum_{j=1}^k{jC_j}}{-\frac{1}{2}k(k+1)+\frac{1}{3}k(2k+1)}.
\end{eqnarray}
The coefficients $a_2(k),$ $b_2(k)$ have analogous formulas with summation from $j=k+1$ to $n.$ Our estimator of the point $l$ in relation (\ref{data}) we define as the number $k$ that minimalizes mentioned sums of distance squares:
\begin{equation}\label{estimator}\hat{l}=
\arg\min_{1\leq k\leq n}{\left[\sum_{j=1}^k{\left(C_j-y_j^1(k)\right)^2}+
\sum_{j=k+1}^n{\left(C_j-y_j^2(k)\right)^2}\right]}.
\end{equation}
Let us stress that the proposed estimator $\hat{l}$ is invariant with respect to sample distribution.

We compare the robustness of detecting the critical point  of the underlying sample satisfying relation (\ref{data}) with the method proposed in \cite{bib:Tsay1988} and based on variance ratio statistics given in \eqref{Tsay:VarianceRatio}. Let us remind that variance ratio statistics is intended to detect change point under the assumption of normal distribution of the examined series. The procedure is the following: we simulate $1000$ trajectories of  length $n=1800$ of stochastically independent random variables with the change point placed on $800$ observation. Similar as for visual pre-tests, we consider two cases each consisting of three distributions. Details of examined scenarios are presented above. 

\begin{figure}[th!]

\centering
\includegraphics[height=5.5cm]{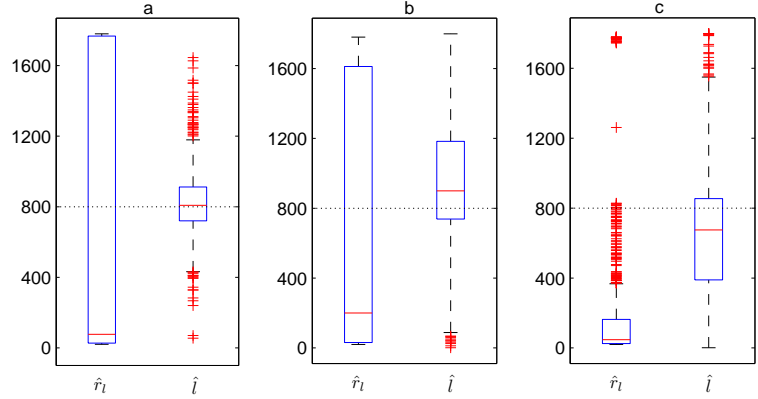}
\caption{Comparison of detection procedure for the critical variance change point for two estimators $\hat{r}_l$ and $\hat{l}$.  Panel a: $\mathcal{N}(0,4),$ $\mathcal{N}(0,4.55),$ panel b: $\mathcal{S}(1.9,0,2,0),$ $\mathcal{S}(1.9,0,2.5,0),$ panel c: $\mathcal{S}(1.8,0,1.2,0),$ $\mathcal{N}(0,2.45).$ }\label{figg1}
\end{figure}

The results for the first scenario, where the critical point is not well visible are presented in Figure \ref{figg1}, where the first boxplot denotes results of $\hat{r}_l$ estimator presented in \eqref{Tsay:VarianceRatio} while the second - is related to $\hat{l}$ estimator defined in  \eqref{estimator}. One clearly sees that detection of the critical point based on the $\hat{l}$ is far more accurate than based on $\hat{r}_l$ even in case of  Gaussian distribution (see panel a in Figure \ref{figg1}).

\begin{figure}[th!]

\centering
\includegraphics[height=5.5cm]{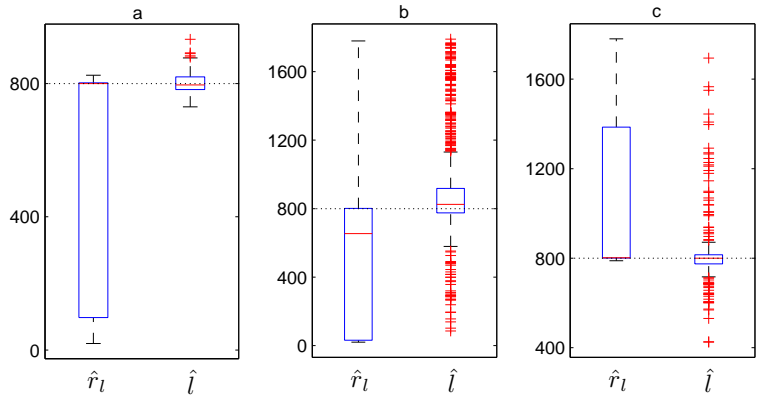}
\caption{Comparison of detection procedure for the critical variance change point for two estimators $\hat{r}_l$ and $\hat{l}$. Panel a: $\mathcal{N}(0,2),$ $\mathcal{N}(0,4),$ panel b: $\mathcal{S}(1.9,0,2,0),$ $\mathcal{S}(1.9,0,4,0),$ panel c: $\mathcal{N}(0,4),$ $\mathcal{S}(1.9,0,1,0).$  }\label{figure2}
\end{figure}

The results for the second scenario with clear critical point are presented in Figure \ref{figure2}. Also in this case one can see that $\hat{l}$ estimator performs better than $\hat{r}_l$.

\section{Statistical test for regime variance}\label{test}
In this Section we introduce the   regime variance test that confirms our assumption of two-regimes behavior given in relation (\ref{data}). It confirms also the preliminary results obtained by using the visual pre-tests presented in the previous Section.  

The procedure is based on the analysis of the empirical second moment of given sample. Let us point  that it can be used for distributions with theoretical second moment but also for this with infinite variance. Even in this case the theoretical second moment exists.  Moreover the test is based on the quantiles that  without assumption of the distribution we can determine on the basis of the empirical distribution function. 

The $\mathcal{H}_0$ hypothesis we define as follows: observed  time series does not satisfy relation (\ref{data}), that means the quantiles of the squared series do not change in time. The hypothesis is satisfied in case of i.i.d. random variables but also in case when distributions of two parts (divided by point $l$) are different but quantiles $q_{\alpha/2}$ and $q_{1-\alpha/2}$ of squared data are on the same level. 

The \textbf{$\mathcal{H}_1$} hypothesis we formulate as: observed time series has at least representation (\ref{data}), i.e. there are at least two regimes of the data for which the appropriate quantiles of squared time series are different. Let us point that the $\mathcal{H}_0$ hypothesis will be rejected when the squared series has more than two regimes. 

The testing of regime variance  is based on the assumption the real data constitute sample of independent variables therefore before testing we have to confirm that given sample constitutes independent data. We propose here to use the simple visual method based on the autocorrelation function (ACF). For independent sample the ACF is close to zero for all lags greater than zero. More information and basic properties of this methodology one can find in \cite{acf}.\\

The procedure of \textbf{regime variance testing} for given time series $X_1,X_2,...,X_n$ proceeds as follows:
\begin{enumerate}

\item Determine the critical point $l$ according to the procedure presented in Section \ref{visual}. Let us emphasize that also under $\mathcal{H}_0$ hypothesis, the $l$ point exists and is between $1$ and $n$.
\item Divide the squared time series into two vectors: $\textbf{W}_1=[X_1^2,...,X_{l}^2]$ and $\textbf{W}_2=[X_{l+1}^2,...,X_{n}^2]$. Find empirical standard deviations $\hat{\sigma}_1$ and $\hat{\sigma}_2$ of $\textbf{W}_1$ and $\textbf{W}_2$, respectively. For simplicity let us assume that $\hat{\sigma}_1<\hat{\sigma}_2$. Let us point that in case of distribution without theoretical second moment, the empirical standard deviation exists and can be calculated on the basis of the observed data. 
\item Construct quantiles from the distribution of squared time series from the vector $\textbf{W}_1$ (for that the empirical standard deviation was smaller), i.e. numbers $q_{\alpha/2}$ and $q_{1-\alpha/2}$ that satisfy the relation
\[P(q_{\alpha/2}<X_i^2<q_{1-\alpha/2})=1-\alpha,~~\mbox{for each $i=1,2,...,l$},\]
where $\alpha$ is a given confidence level. Under the $\mathcal{H}_0$ hypothesis without the assumption of the distribution, the appropriate quantiles we can determine on the basis on the empirical cumulative distribution function. Because $X_{l+1}^2,X_2^2,...,X_n^2$ are independent therefore the statistics $B$ has Binomial distribution with parameters $n-l$ and $p=1-\alpha$. Therefore the p-value of the test we calculate as $P(Z<B)$, where $Z$ has Binomial distribution with $(n-l,p)$ parameters.
\item If the calculated p-value is greater than the $\alpha$ parameter, then we accept the $\mathcal{H}_0$ hypothesis. Otherwise if the calculated p-value is smaller than the $\alpha$ parameter, then we reject the $\mathcal{H}_0$ hypothesis and accept $\mathcal{H}_1$.
\end{enumerate}

The complementary part of this Section is the simulation examination of the proposed estimator (\ref{estimator}) and variance regime test described above. First we check the committed error of the first order for our test, i.e. the rejecting a true $\mathcal{H}_0$ hypothesis. For this purpose we generate 1000 trajectories of  length $1800$ of stochastically independent random variables for each of three cases:
\begin{itemize}
\item{the Gaussian case with $\mathcal{N}(0,2)$ distribution,}
\item{the L\'evy--stable case with $\mathcal{S}(1.8,0,1,0)$ distribution,}
\item{the Gaussian--L\'evy--stable case with $\mathcal{N}(0,1)$ and $\mathcal{S}(1.9,0,1,0)$ distribution for each half of the sample, randomly permuted.}
\end{itemize}
In our simulations we always assume the significance level $\alpha=0.05$ and the unknown distribution of samples. Therefore in the testing procedure the empirical quantiles are applied. We note that the first two cases (pure Gaussian and L\'evy--stable) are the special simplified versions of $\mathcal{H}_0,$ i.e. i.i.d. data. Obviously the constancy of theoretical quantiles $q_{\alpha/2}$ and $q_{1-\alpha/2}$  implies the closeness between the empirical versions computed in testing algorithm. The Gaussian--L\'evy--stable case concerns  two different distributions of data changing dynamically (randomly permuted) on the time domain. Therefore it is contrast to the $\mathcal{H}_1$ hypothesis where different distributions are concentrated in two disjoint time intervals.

\begin{figure}[th!]

\includegraphics[height=5.5cm]{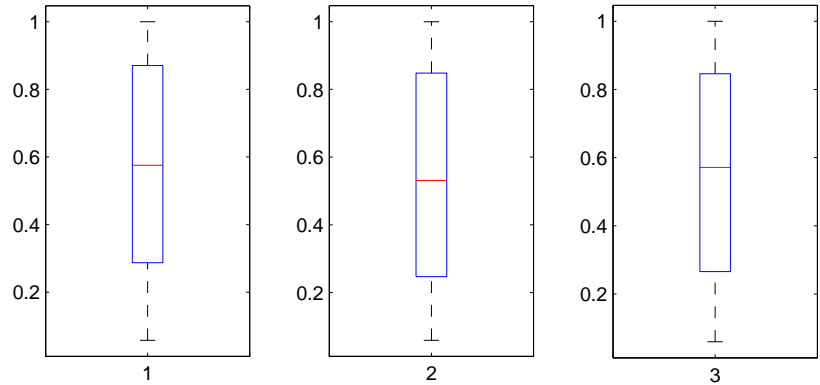}
\caption{The boxplots of p-values corresponding to the correct accepting of true $\mathcal{H}_0:$  1) $\mathcal{N}(0,2),$ 
2) $\mathcal{S}(1.8,0,1,0),$ 3) permuted $\mathcal{N}(0,1),$ $\mathcal{S}(1.9,0,1,0).$}\label{box1}
\end{figure}

The results of conducted simulations we present in Table \ref{tab1}. For testing procedure we apply the sample mean value of obtained estimators $\hat{l}$ from each generated sample. That mean value of $\hat{l}$ is 881.37, 916.34 and 104.31 for each of three considered cases, respectively. They are close to the half of the sample length, which is quite intuitive for data satisfying $\mathcal{H}_0$. We see that extremely more times the test correctly does not reject the true null hypothesis $\mathcal{H}_0$ and the error of the first order is strongly rare, see the $\mathcal{H}_1$--column. Moreover, the p-values corresponding to the acceptance of $\mathcal{H}_0$ are rightly higher than significance level $\alpha=0.05,$ see Figure \ref{box1}. The column of Table \ref{tab1} with p-value contains the mean of such p-values. Moreover in the $\mathcal{H}_0$--column and $\mathcal{H}_1$--column we present the numbers of correct accepting and incorrect rejecting $\mathcal{H}_0$, respectively.
\begin{table}[bh!]
\label{tab1}
\centering
\begin{tabular}{|c|c|c|c|}
\hline
 Distribution of samples& $\mathcal{H}_0$  &    p-value    &   $\mathcal{H}_1$ \\\hline
\hline
 $\mathcal{N}(0,2)$  &866  &  0.5623 &   134    \\\hline

 $\mathcal{S}(1.8,0,1,0)$  &865 &   0.5349  &  135   \\\hline

$\mathcal{N}(0,2),$ $\mathcal{S}(1.9,0,1,0)$ &889 &   0.5621  & 111   \\\hline
\end{tabular}
\caption{Numbers of correct accepting (with mean p-value) and incorrect rejecting the true $\mathcal{H}_0.$}\label{tab1}
\end{table}
\newpage
Our next task is to explore the statistical power of the examined test. This is equivalent issue to investigation of committing the error of the second order, i.e.  accepting a false $\mathcal{H}_0$ hypothesis. In order to calculate the error of the second order, we simulate 1000 trajectories of  length $1800$ of stochastically independent random variables for each of three cases from the first scenario described in Section \ref{visual} satisfying the $\mathcal{H}_1$ hypothesis.

\begin{figure}[h!]
\includegraphics[height=5.5cm]{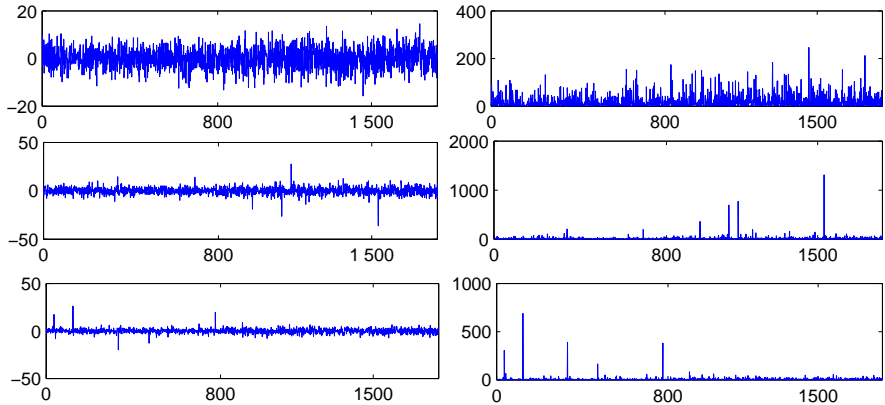}
\caption{The exemplary samples (left panels) and squared samples (right panels) for three considered cases of $\mathcal{H}_1:$ 1) $\mathcal{N}(0,4),$ $\mathcal{N}(0,4.55),$
2) $\mathcal{S}(1.9,0,2,0),$ $\mathcal{S}(1.9,0,2.5,0),$ 3) $\mathcal{S}(1.8,0,1.2,0),$ $\mathcal{N}(0,2.45).$ }\label{exemtraj}
\end{figure}

In all three cases the differences of distribution parameters are quite small and the $\mathcal{H}_1$ hypothesis statement can be invisible from the data or its squares, see Figure \ref{exemtraj}. This means that we check the efficiency of proposed test in a very sophisticated cases. 

\begin{figure}[th!]
\includegraphics[height=5.5cm]{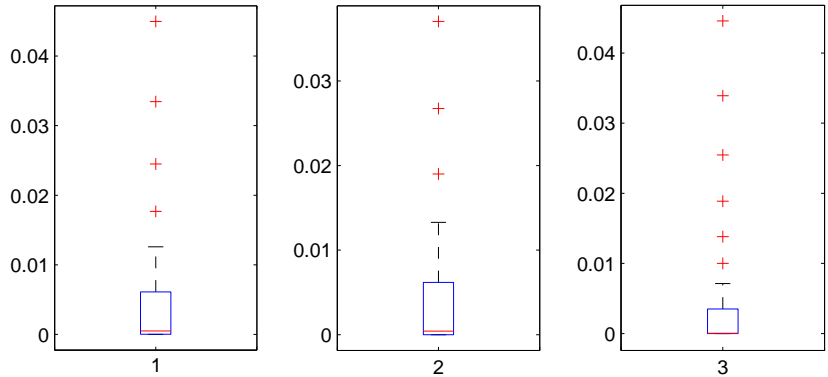}
\caption{The boxplots of p-values corresponding to the correct rejecting of false $\mathcal{H}_0:$ 1) $\mathcal{N}(0,4),$ $\mathcal{N}(0,4.55),$
2) $\mathcal{S}(1.9,0,2,0),$ $\mathcal{S}(1.9,0,2.5,0),$ 3) $\mathcal{S}(1.8,0,1.2,0),$ $\mathcal{N}(0,2.45).$ }\label{box2}
\end{figure}

We apply the estimator (\ref{estimator}) and adopt the  regime variance test assuming the unknown data distribution. The results of conducted simulations with significant level $\alpha=0.05$ we present in  Table \ref{tab2}. For testing procedure we apply the sample mean value of obtained estimators $\hat{l}$ from each generated sample. That mean value of $\hat{l}$ is 822.28, 943.72 and 646.42 for each of three considered cases from the first scenario, respectively. We see that more times the test correctly reject the false null hypothesis $\mathcal{H}_0$ and the error of the second order is rare, see the $\mathcal{H}_0$--column. 
The worst result we obtain in the third case with different distributions. However the p-values corresponding to the rejection of $\mathcal{H}_0$ are rightly lower than significance level $\alpha=0.05,$ see Figure \ref{box2}. The column of Table \ref{tab2} with p-value contains the mean of such p-values. Moreover in the $\mathcal{H}_0$--column and $\mathcal{H}_1$--column we present the numbers of incorrect accepting $\mathcal{H}_0$ and correct accepting $\mathcal{H}_1$ (the power of the test), respectively. We also strongly stress that from the construction of the studied test the rejection of $\mathcal{H}_0$ hypothesis is equivalent to the acceptance of $\mathcal{H}_1.$ In other words, the rejection of $\mathcal{H}_0$ is only possible when $\mathcal{H}_1$ is true or in the case of first order error.
\begin{table}[bh!]
\label{tab2}
\centering
\begin{tabular}{|c|c|c|c|}
\hline
 Distribution of samples& $\mathcal{H}_1$  &    p-value    &   $\mathcal{H}_0$ \\\hline
\hline
 $\mathcal{N}(0,4),$ $\mathcal{N}(0,4.55)$  &759  &  0.0061 &   241    \\\hline

$\mathcal{S}(1.9,0,2,0),$ $\mathcal{S}(1.9,0,2.5 ,0)$  &758 &   0.0054  &  242   \\\hline

$\mathcal{S}(1.8,0,1.2,0),$ $\mathcal{N}(0,2.45)$ &652 &   0.0044  &  348   \\\hline
\end{tabular}
\caption{Numbers of correct rejecting (with mean p-value) and incorrect accepting the false $\mathcal{H}_0.$}\label{tab2}
\end{table}
\section{Plasma data analysis}
\begin{figure}[htb]
\centering \includegraphics[height=6cm]{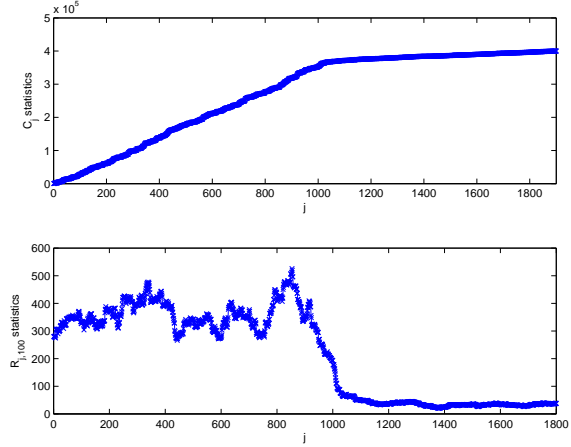}
\caption{The  visual pre-tests for regimes variance of time series that presents increments of floating
potential fluctuations  of turbulent laboratory plasma for the  small torus radial position $r=37$ cm. }\label{visual_time_series_1}
\end{figure}

In this Section we analyze the real data sets presented in Figure \ref{time_series_1} and \ref{time_series_2} by using the tests for regime variance described in Sections \ref{visual} and \ref{test}. In Figure \ref{visual_time_series_1}  we demonstrate results of the visual pre-tests  for increments of floating potential fluctuations  of turbulent laboratory plasma for the  small torus radial position $r=37$ cm (corresponding to Figure \ref{time_series_1}). 

\begin{figure}[htb]
\centering \includegraphics[height=3cm]{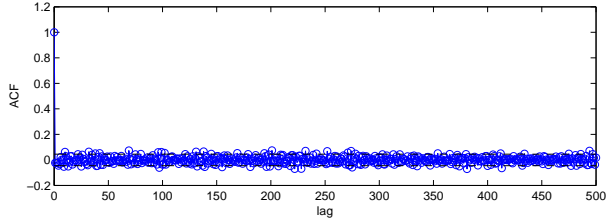}
\caption{ACF of the time series that presents increments of floating
potential fluctuations  of turbulent laboratory plasma for the  small torus radial position $r=37$ cm. Such behavior of autocorrelation function suggests that the data are independent.}\label{acf_time_series_1}
\end{figure}

As we observe, the visual pre-tests indicate at the behavior formulated in (\ref{data}). Moreover we can also determine the critical point $l$, that divides the time series into two independent samples with appropriate statistical properties that do not change in time.  We estimate the  point by using the procedure described in Section \ref{visual} and get $1055$. In the next step of our analysis we test the $\mathcal{H}_0$, i.e. the hypothesis that  the squared time series has quantiles that do not change in time. According to the procedure presented in Section \ref{test} first we confirm  independence by using ACF function, see Figure \ref{acf_time_series_1}. 

The regime variance test  confirms that the examined data set has at least two regimes, i.e. it has representation (\ref{data}). This is related to the fact that with confidence level $\alpha=0.05$,  the obtained p-value is equal to $0.0425$ (we reject $\mathcal{H}_0$).  Because we have estimated the critical point $l$, that divides the analyzed time series into two parts, therefore we can examine if the separate vectors can be considered as independent samples with the same appropriate quantiles of squared data. In order to do this, we use the regime variance test once again for samples $X_1,...,X_{1055}$ and $X_{1056},...X_{1900}$. For the first time series,  the test returns p-value on the level $0.5967$, that indicates we can assume the squared data have appropriate quantiles that do not change in time. If we test the second part of the data set, namely observations from $1056$ to $1900$, we get p-value equal to $0.9829$, therefore also for this vector we can conclude that appropriate statistical properties do not change. Moreover if we assume the data from two considered parts constitute i.i.d. samples (that is one of the possibility when $\mathcal{H}_0$ hypothesis is satisfied), we can  test the distributions. By using tests based on the empirical cumulative distribution function completely described in \cite{nasza}, we conclude the observations $X_1,...,X_{1055}$ come from  L\'{e}vy--stable distribution with stable parameter equal to $1.76$ and $\sigma=12.14$,  while the data $X_{1056},...,X_{1900}$ have  L\'{e}vy--stable distribution with parameters $\alpha=1.91$ and $\sigma=4.2$. For both samples we use the McCulloch's estimation method, \cite{reg}.

\begin{figure}[htb]
\centering \includegraphics[height=6cm]{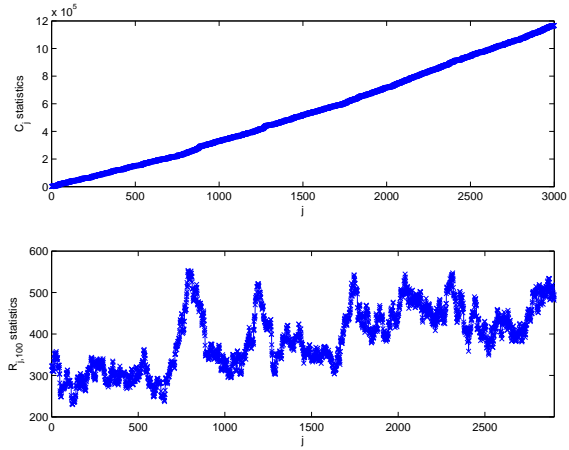}
\caption{The  visual pre-tests for regimes variance of time series that presents increments of floating
potential fluctuations  of turbulent laboratory plasma for the  small torus radial position $r=36$ cm. }\label{visual_time_series_2}
\end{figure}

\begin{figure}[htb]
\centering \includegraphics[height=3cm]{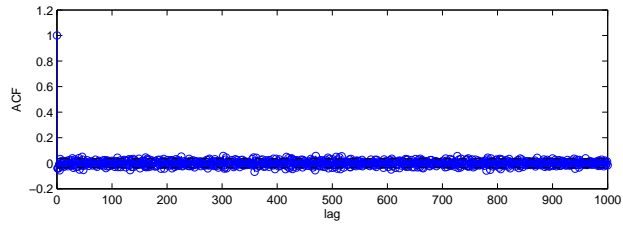}
\caption{ACF of the time series that presents increments of floating
potential fluctuations  of turbulent laboratory plasma for the  small torus radial position $r=36$ cm. Such behavior of autocorrelation function suggests that the data are independent.}\label{acf_time_series_2}
\end{figure}

As we have mentioned in Section \ref{moti}, the testing procedure we can use  also for data for which the critical point $l$  is not so visible as in the previous case, see Figure \ref{time_series_2}. In Figure \ref{visual_time_series_2} we present results of the visual pre-tests described in Section \ref{visual} for data that describes increments of floating
potential fluctuations  of turbulent laboratory plasma for the  small torus radial position $r=36$ cm. As we observe, on the basis of the behavior of $C_j$ and $R_{j,k}$ statistics defined in (\ref{Ck}) and  (\ref{R}) respectively we can not conclude that the data set exhibits behavior described in (\ref{data}). But the procedure of estimating the critical point returns $1763$. 
According to the scheme of regime variance testing presented in Section \ref{test}, in the first step we confirm independence of the time series. The plot of ACF is presented in Figure \ref{acf_time_series_2}. 

Next we can test if the hypothesis $\mathcal{H}_0$ is satisfied for time series presented in Figure \ref{time_series_2}. The obtained p-value equal to $0.0011$ indicates the data has at lest two regimes with different statistical properties. Similar, as for the first data set, we divide the time series into two separate vectors and test if we can consider them as samples for which the characteristics do not change with respect to time. For the first  part, namely data from $1$ to $1763$ we get p-value on the level $0.593$, while for the second vector (i.e. observations form $1764$ to $3000$) the p-vale is equal to $0.591$. These results indicate that  two considered  vectors do not satisfy relation (\ref{data}) and can be considered as i.i.d. samples. Under this assumption we test the distributions and obtain that the two considered parts come from L\'{e}vy--stable distribution. For the first vector we obtain following estimates of the parameters:  $\alpha=1.9484$ and  $\sigma=12.9505$, while  estimated values of parameters for the vector containing observations $X_{1764},...,X_{3000}$ are:  $\alpha=1.7983$ and $\sigma=14.1099$.
\section*{Conclusions}\label{conclusions}
This paper is devoted to analysis of time series that exhibit two-regimes behavior. We have introduced the new estimation procedure  for recognition the critical point that divides the observed time series into two regimes with different statistical properties expressed in the language of the quantiles for squared data (Section \ref{visual}). We have developed also three tests  that can confirm our assumption of two-regimes behavior (Sections \ref{visual} and \ref{test}). The universality of the presented methodology comes from the fact that it does not assume the distribution of examined time series therefore it can by applied to rich class of real data sets.  The theoretical results we have  illustrated by using the simulated time series and analysis of two real data sets related to turbulent laboratory plasma. 

\section*{Acknowledgment}
The JG and GS would like to acknowledge a partial support of the Fellowship co-financed by European Union within European Social Fund.

\end{document}